\documentclass{aa}  
\usepackage[utf8]{inputenc}
\usepackage{graphicx}
\usepackage{txfonts}
\usepackage{threeparttable}
\usepackage{xcolor}
\usepackage[normalem]{ulem}
\begin{document}

   \title{Formation of nitriles and isonitriles by the heavy-ion irradiation of propionitrile in N$_2$-rich astrophysical ices}





   \author{A.L.F. de Barros
          \inst{1}
          \and
           D. Dubois \inst{1}
         \and
       R. Sreeja\inst{1}
\and
T. Nguyen Trung\inst{1}
\and
T. Sbera\inst{2}
\and
H. Rothard\inst{1}
\and A. Domaracka\inst{1}
        }

   \institute{$^1$Centre de Recherche sur les Ions, les Matériaux et la Photonique (CIMAP), Université Caen Normandie, ENSICAEN, CNRS, CEA, Normandie Univ., UMR 6252, Caen, F-14000, France\\
   $^2$Departamento de Física, Centro Federal de Educação Tecnológica Celso Suckow da Fonseca (CEFET/RJ), Av. Maracanã 229, Rio de Janeiro, 20271-110, RJ, Brazil\\
              }

   \date{Received July 07, 2026}

 \abstract
{Nitriles are key nitrogen-bearing organic molecules in dense clouds, star-forming regions, and nitrogen-rich icy environments. Understanding their stability and chemical evolution under energetic processing is essential for understanding the formation of complex organic species in astrophysical ices.}
{We investigate the radiolytic processing of propionitrile (CH$_3$CH$_2$CN, hereafter referred to as PCN) in a  nitrogen-rich ice matrix and evaluate the formation of nitriles, isonitriles, hydrocarbons, and other nitrogen-bearing products induced by swift heavy ions.}
{A PCN:N$_2$ ice mixture with an approximate molecular ratio of 1:10 was deposited at 10 K and irradiated with 40 MeV $^{40}$Ar$^{9+}$ ions up to a fluence of $1\times10^{13}$ ions cm$^{-2}$. The chemical evolution was monitored in situ by Fourier-transform infrared spectroscopy. Destruction and formation cross sections, as well as radiation chemical yields, were derived from the fluence dependence of selected infrared bands.}
{Ion irradiation efficiently destroys PCN and produces a rich inventory of daughter species. The products include: nitriles and isonitriles such as HCN, HCNN, HC$_3$N, CH$_3$CN, CH$_3$C$_3$N, CH$_3$CHCNH, CH$_2$CHCN, NCCN/C$_2$N$_2$, CN, and C$_2$N; nitrogen-bearing species such as CH$_2$NH, CH$_3$NH$_2$, CH$_3$N$_3$, and N$_3^{-}$; and hydrocarbons, including CH$_4$, C$_2$H$_2$, C$_2$H$_4$, C$_2$H$_6$, and C$_4$H$_4$. The derived cross sections indicate that CN-bearing fragments and hydrocarbons are among the most efficiently formed products, demonstrating that the C$\equiv$N group is efficiently preserved and that extensive carbon-chain reorganization also occurs.}
{The results demonstrate that the energetic processing of PCN in N$_2$-rich ices provides an efficient pathway to molecular complexity under conditions relevant to dense interstellar clouds, protostellar environments, and nitrogen-rich outer Solar System surfaces.}


   \maketitle
%

\section{Introduction}

Nitrogen-bearing organic molecules are widespread throughout the interstellar medium (ISM), star-forming regions, protoplanetary disks, and Solar System bodies. Among them, nitriles constitute one of the most abundant and chemically diverse families, including simple molecules such as HCN and CH$_3$CN as well as more complex species such as HC$_3$N, HC$_5$N, CH$_2$CHCN, and CH$_3$CH$_2$CN \citep{Belloche2022,Daly2013}. Because of the exceptional stability of the C$\equiv$N functional group, these molecules survive under a wide range of astrophysical conditions and participate in reaction networks, leading to increasingly complex organic compounds.

Many nitriles are expected to accrete onto icy grain mantles together with abundant volatile species such as H$_2$O, CO, CO$_2$, CH$_3$OH, NH$_3$, and N$_2$. Within these ices, continuous exposure to ultraviolet photons, secondary electrons, and energetic cosmic rays drives molecular dissociation, ionization, and radical recombination, promoting the synthesis of new organic molecules \citep{Hudson2004,Moore2010,Krim2019}. Laboratory simulations have shown that the energetic processing of nitrile-rich ices efficiently produces hydrocarbons, cyanopolyynes, isonitriles, and other nitrogen-bearing species, highlighting the important role of energetic chemistry in the evolution of astrophysical ices. 

Among the astrophysical nitriles, propionitrile --- CH$_3$CH$_2$CN (PCN hereafter), also known as ethyl cyanide is of particular interest because it represents the next member of the homologous series after acetonitrile and has been detected toward several hot molecular cores, including Sagittarius B2(N), where it is associated with complex organic chemistry \citep{Daly2013,Nazari2024}. Molecules of similar complexity have also been proposed as key constituents of Titan's atmosphere and other nitrogen-rich planetary environments in the Solar System \citep{Czaplinski2025,Nixon2024,Coustenis2010}. Despite its astrophysical relevance, however, the radiation chemistry of condensed PCN remains poorly understood.

Infrared spectra of solid PCN and related nitriles have been reported since the pioneering studies of \citet{Heise1981} and \citet{Del96}, providing the spectroscopic basis for laboratory astrochemistry. More recently, ultraviolet photolysis and ion irradiation experiments on CH$_3$CN-containing ices have revealed an efficient formation of HCN, HC$_3$N, cyanogen, unsaturated nitriles, hydrocarbons, and other nitrogen-bearing molecules \citep{Hudson2004,Coupeaud2006,Moore2010,Danger2013,Krim2019,Abdulgalil2013,Borengasser2026RadiolysisChemistry}. In contrast, comparable studies involving PCN are still scarce, particularly under swift heavy-ion irradiation, which constitutes one of the best laboratory analogs of heavy cosmic rays because of its high electronic stopping power and dense energy deposition.

\begin{figure*}[t]
\centering
\includegraphics[width=0.98\textwidth]{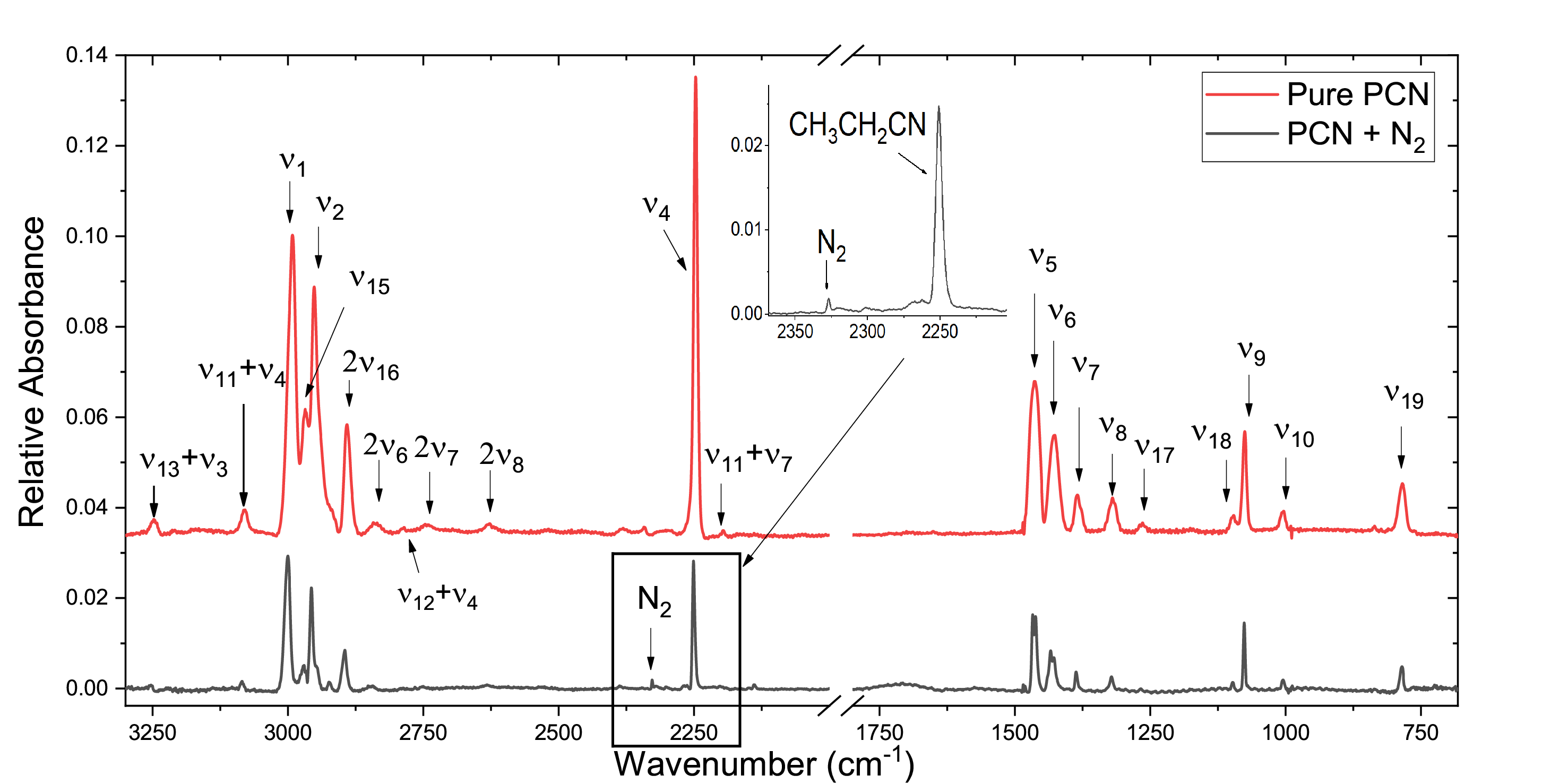}
\caption{Infrared spectra of pure PCN (upper spectrum) and the PCN:N$_2$ (1:10) ice mixture (lower spectrum) deposited at 10 K before irradiation. The main vibrational bands of PCN and the matrix-induced absorption of solid N$_2$ are indicated.}
\label{fig1}
\end{figure*}

In this work, we investigate the radiolytic processing of a pre-mixed PCN:N$_2$ ice with a molecular ratio of approximately 1:10 deposited at 10 K and irradiated with 40 MeV $^{40}$Ar$^{9+}$ ions. The chemical evolution of the ice was monitored in situ by Fourier-transform infrared (FTIR) spectroscopy, which enabled the identification of newly formed species and the determination of their destruction and formation cross sections, together with the corresponding radiation chemical yields. Particular emphasis is placed on the formation of nitriles, nitrogen-bearing molecules, and hydrocarbons produced through the energetic processing of PCN in a molecular nitrogen matrix, providing new constraints on the chemistry of nitrogen-rich astrophysical ices exposed to cosmic-ray irradiation.
Figure~\ref{fig1} shows the infrared spectra of pure PCN and of the PCN:N$_2$ ice mixture deposited at 10 K before irradiation, providing the spectroscopic reference for the subsequent analysis.

\section{Experimental procedures}
\label{sec:exp1}

The irradiation experiments were carried out at the Centre de recherche sur les Ions, les MAtériaux et la Photonique (CIMAP) at Centre Interdisciplinaire de Recherches Ions Lasers (CIRIL) laboratory using the Irradiation Sud beamline of the Grand Accélérateur National d'Ions Lourds (GANIL; Caen, France). The experimental setup has been described in detail elsewhere, and only the parameters relevant to the present work are summarized here \citep{aug18,sep21,mej13,bar25,deBarros2026}.
A pre-mixed gaseous PCN:N$_2$ mixture with an approximate molecular ratio of 1:10 was deposited onto an infrared-transparent ZnSe substrate maintained at 10 K in an ultra-high-vacuum chamber ($P<5\times10^{-10}$ mbar). Propionitrile (99.9\% purity) was purified by several freeze--pump--thaw cycles prior to deposition to remove volatile contaminants. Ultra-high-purity N$_2$ ($\geq99.9999\%$, Air Liquide) was used. The resulting ice corresponds to an atomic composition of C = 10.34\%, H = 17.24\%, and N = 72.42\%.

The ice was irradiated with 40 MeV $^{40}$Ar$^{9+}$ ions using an ion flux of approximately $1.3\times10^{9}$ ions cm$^{-2}$ s$^{-1}$ up to a maximum fluence of $1\times10^{13}$ ions cm$^{-2}$. Electronic and nuclear stopping powers were calculated with the SRIM code \citep{ziegler2010srim}, yielding $S_e = 2.62\times10^{3}$ keV $\mu$m$^{-1}$ and $S_n = 4.39$ keV $\mu$m$^{-1}$, respectively, confirming that the energy deposition is dominated by electronic interactions. 
Infrared spectra were acquired in situ in transmission mode using a FTIR spectrometer between 5000 and 600 cm$^{-1}$ with a spectral resolution of 1 cm$^{-1}$. Each spectrum corresponds to the average of 300 scans after subtraction of the background spectrum recorded before ice deposition \citep{bar24}.

Column densities were calculated from the Beer--Lambert relation:

\begin{equation}
N=\frac{1}{A_v}\int\tau(\nu)\,d\nu,
\end{equation}

\noindent where $N$ is the column density, $\tau(\nu)$ is the optical depth, and $A_v$ is the band strength of the selected vibrational mode. For PCN, the C$\equiv$N stretching band at 2247 cm$^{-1}$ was used with $A_v = 3.3\times10^{-18}$ cm molecule$^{-1}$ calculated from the integrated absorption coefficient reported by \citep{Del96}, resulting in an initial column density of $N_0(\mathrm{PCN}) = 5.4\times10^{17}$ molecules cm$^{-2}$. Molecular nitrogen was monitored through the weak matrix-induced absorption at 2327 cm$^{-1}$ using  $A_v = 1.8\times10^{-22}$ cm molecule$^{-1}$ \citep{Bernstein1999}, giving an initial column density of $N_0(\mathrm{N_2}) \approx 5.0\times10^{18}$ molecules cm$^{-2}$. The corresponding band assignments for the precursor are summarized in Table \ref{tab:table1}. Column densities of the daughter species were determined using literature band strengths whenever available; the adopted assignments and $A_v$ values are listed in Table~\ref{tab:table2}.

\begin{figure*}[t]
\centering
\includegraphics[width=0.98\textwidth]{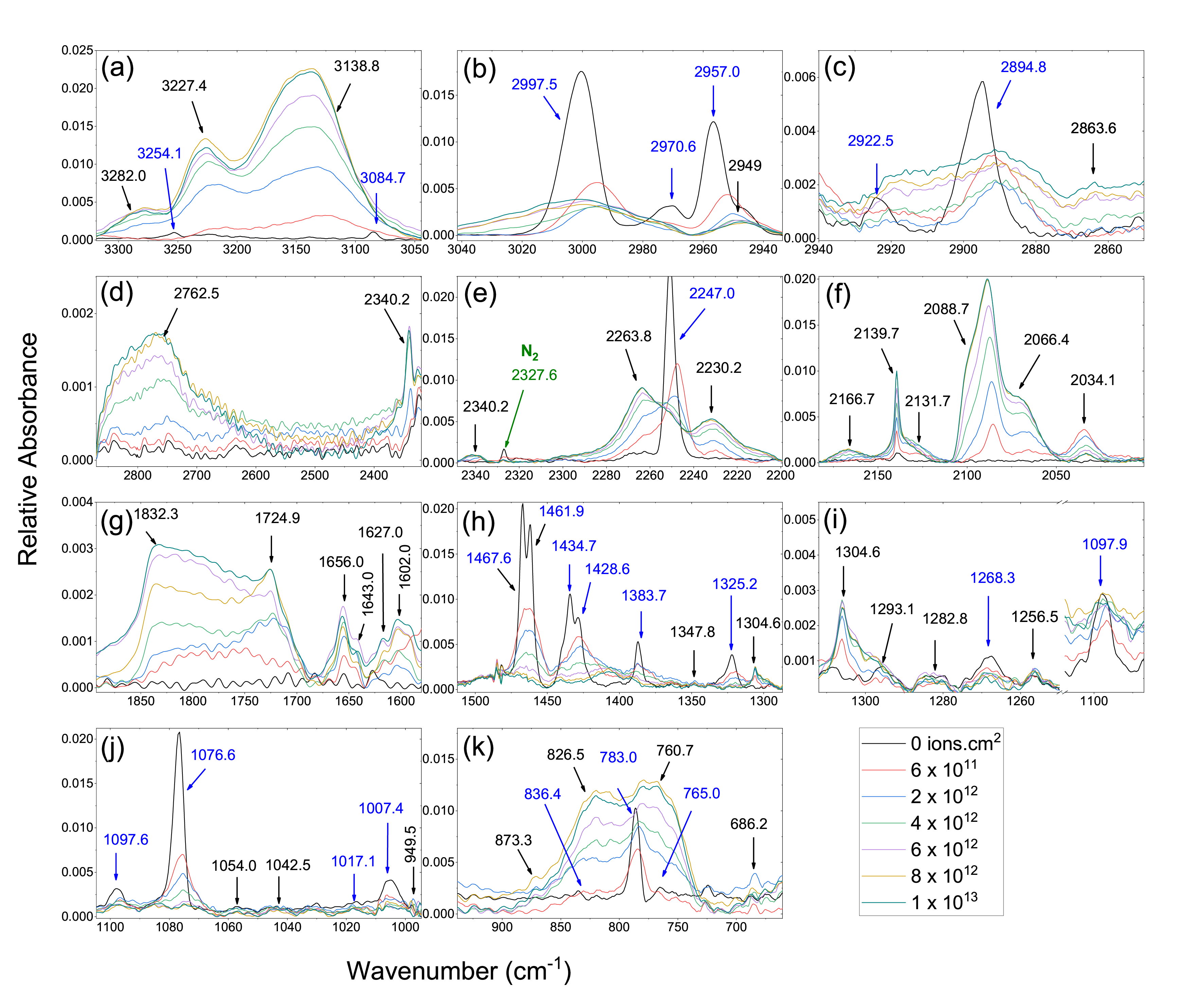}
\caption{Infrared spectra of irradiated PCN:N$_2$ ice at 10 K. The different panels display selected spectral regions used to identify precursor and product species. Assignments corresponding to the precursor PCN are marked in blue, newly formed bands in black, and N$_2$-related features in green. Spectra are shown for increasing ion fluences from 0 to 1 $\times$ 10$^{13}$ ions cm$^{-2}$.}
\label{fig2}
\end{figure*}




\begin{figure*}[t]
\centering
\includegraphics[width=\textwidth]{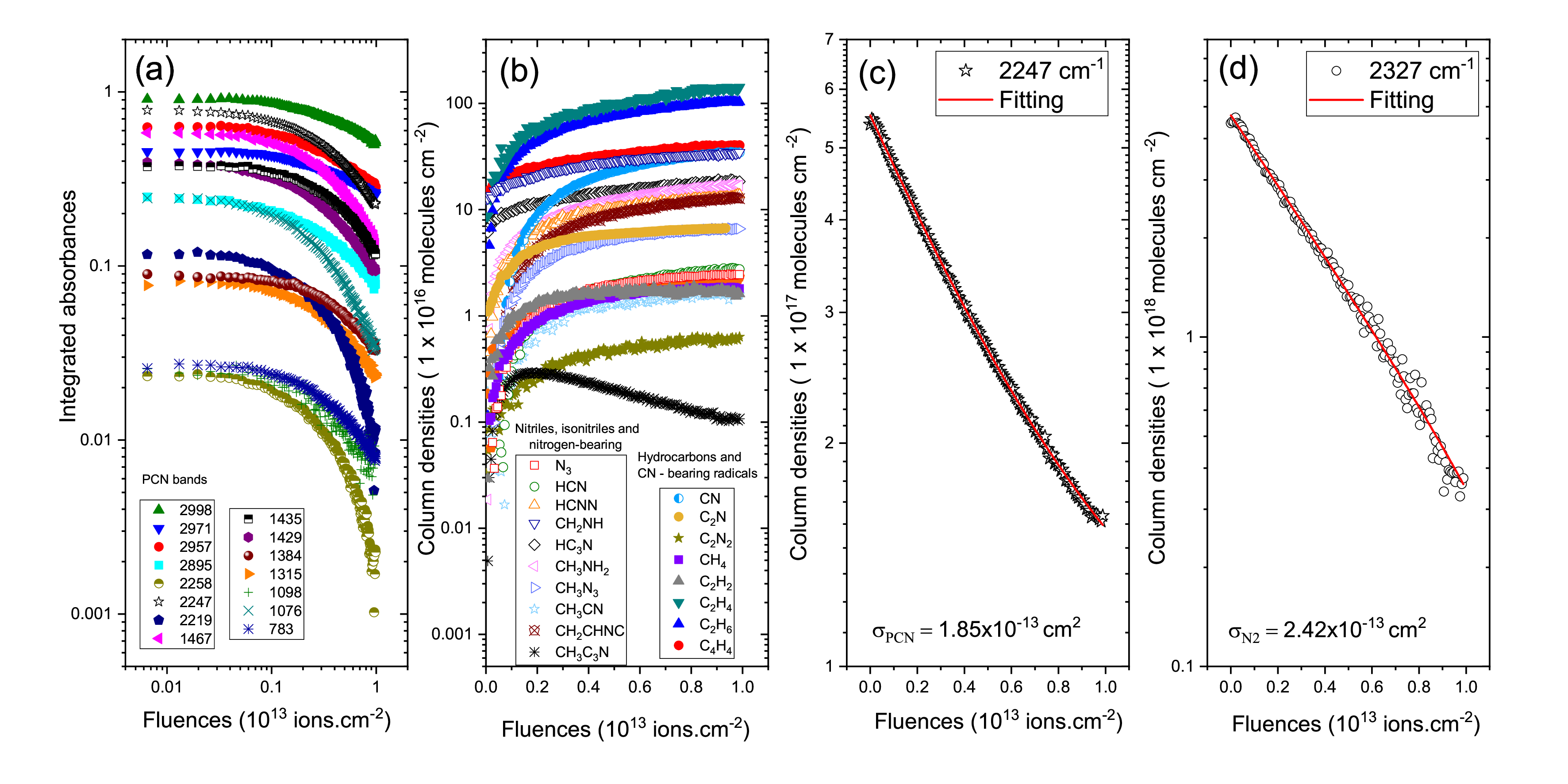}
\caption{(a) Integrated absorbance areas of the main infrared bands assigned to PCN as a function of ion fluence. (b) Column density evolution of the identified daughter species, grouped according to chemical families. (c) Exponential fit of the PCN column density derived from the 2247 cm$^{-1}$ C$\equiv$N stretching band. (d) Exponential fit of the matrix-induced N$_2$ band at 2327 cm$^{-1}$. The two fits were used to determine the effective disappearance cross sections; for details, see Sect. 3.3.}
\label{fig3}
\end{figure*}

\section{Results and discussion}

\subsection{Infrared spectrum of pristine PCN:N$_2$ ice}

The infrared spectrum of the pristine ice provides the spectroscopic reference for evaluating the chemical modifications induced by ion irradiation. Figure~\ref{fig1} compares the spectra of pure PCN deposited at 10 K with that of the PCN:N$_2$ mixture used in the present work. The overall spectral profile remains essentially unchanged after dilution in molecular nitrogen, indicating that the vibrational fingerprint of PCN is preserved within the N$_2$ matrix.

Only minor frequency shifts and slight band broadening are observed for a few vibrational modes. These variations arise from matrix-isolation effects associated with changes in the local molecular environment and have previously been reported for nitriles isolated in cryogenic nitrogen matrices \citep{Bohn1994,Moore2010}. No additional absorption features are detected before irradiation, confirming that the deposition process does not induce measurable chemical modifications.

The principal absorption bands of the pristine PCN:N$_2$ ice are shown in Fig.~\ref{fig2}e. The strongest absorption at 2247 cm$^{-1}$ corresponds to the C$\equiv$N stretching mode of PCN and was used to determine the precursor column density throughout the irradiation experiment. The remaining bands are assigned to CH$_3$ and CH$_2$ stretching, deformation, rocking, wagging, skeletal, and C--N vibrational modes, in good agreement with previous infrared studies of condensed PCN \citep{Del96,nnamvondo2019}. A weak absorption centered near 2327 cm$^{-1}$ is assigned to matrix-induced molecular nitrogen. Although N$_2$ is infrared inactive in the gas phase, weak absorptions become allowed in the condensed phase owing to symmetry breaking induced by neighboring molecules \citep{Bohn1994,Khanna2005CondensedData}. This feature confirms the incorporation of molecular nitrogen into the deposited ice.

\subsection{Evolution of the infrared spectra during irradiation}

Figure~\ref{fig2} presents the infrared spectra of the PCN:N$_2$ ice recorded at several ion fluences. As irradiation proceeds, the precursor bands progressively decrease in intensity, while numerous new absorption features appear over the entire spectral range, indicating the simultaneous destruction of the precursor molecules and the formation of radiolysis products. The labeled absorption bands provide the spectroscopic reference for identifying both the precursor and the daughter species, and their assignments are summarized in Table~\ref{tab:table1}.

The most pronounced spectral changes occur in the C$\equiv$N stretching region (2350--2000 cm$^{-1}$), where the depletion of the PCN band at 2247 cm$^{-1}$ is accompanied by the appearance of several new absorptions attributed to nitriles and isonitriles, including HCN, HC$_3$N, CH$_3$CN, CH$_2$CHCN, CH$_3$C$_3$N, cyanogen (NCCN/C$_2$N$_2$), CH$_3$NC, HCNN, and tentative ketenimine-like species [(CH$_3$)HCCNH / H$_2$CCNH]. Additional bands assigned to the CN and C$_2$N radicals indicate extensive fragmentation of the precursor and the production of highly reactive intermediates that subsequently participate in radical recombination reactions \citep{Wu2013}.

New absorption bands are also observed outside the C$\equiv$N region, revealing the formation of nitrogen-bearing molecules such as CH$_2$NH, CH$_3$NH$_2$, CH$_3$N$_3$, and N$_3^-$, together with hydrocarbons including CH$_4$, C$_2$H$_2$, C$_2$H$_4$, C$_2$H$_6$, and C$_4$H$_4$. The simultaneous formation of these species demonstrates that energetic processing of the mixed PCN:N$_2$ ice promotes an extensive reaction network involving molecular fragmentation, hydrogen abstraction, dehydrogenation, radical recombination, isomerization, and carbon-chain growth. In particular, the formation of nitriles such as HC$_3$N, CH$_3$CN, CH$_2$CHCN, and CH$_3$C$_3$N indicates that fragmentation of the PCN molecule is frequently accompanied by hydrogen rearrangement and carbon-chain growth while preserving the nitrile functionality. Similar reaction pathways have been reported for energetic processing of nitrile-containing astrophysical ice analogs \citep{Hudson2004,Abdulgalil2013,Danger2013,Borengasser2026RadiolysisChemistry}.

Although cyclization reactions leading to aromatic nitrogen heterocycles such as pyridine are chemically plausible during the energetic processing of unsaturated nitriles, no evidence of pyridine formation was obtained under the present experimental conditions. Simultaneous monitoring of the gas phase by the in situ quadrupole mass spectrometer (QMS) showed intense signals at m/z = 28 during deposition and irradiation, mainly associated with molecular nitrogen and possible PCN fragmentation. Signals at m/z = 54, attributed to PCN-related fragments, were mainly observed during deposition, while no significant signal at m/z = 79 was detected above the background during irradiation. Since m/z = 79 corresponds to the molecular ion of pyridine, the QMS measurements provide additional evidence that pyridine was not formed in detectable amounts under the present experimental conditions.  The complete infrared assignments of the newly formed species are summarized in Table~\ref{tab:table1}.


\begin{figure*}[t]
\centering
\includegraphics[width=\textwidth]{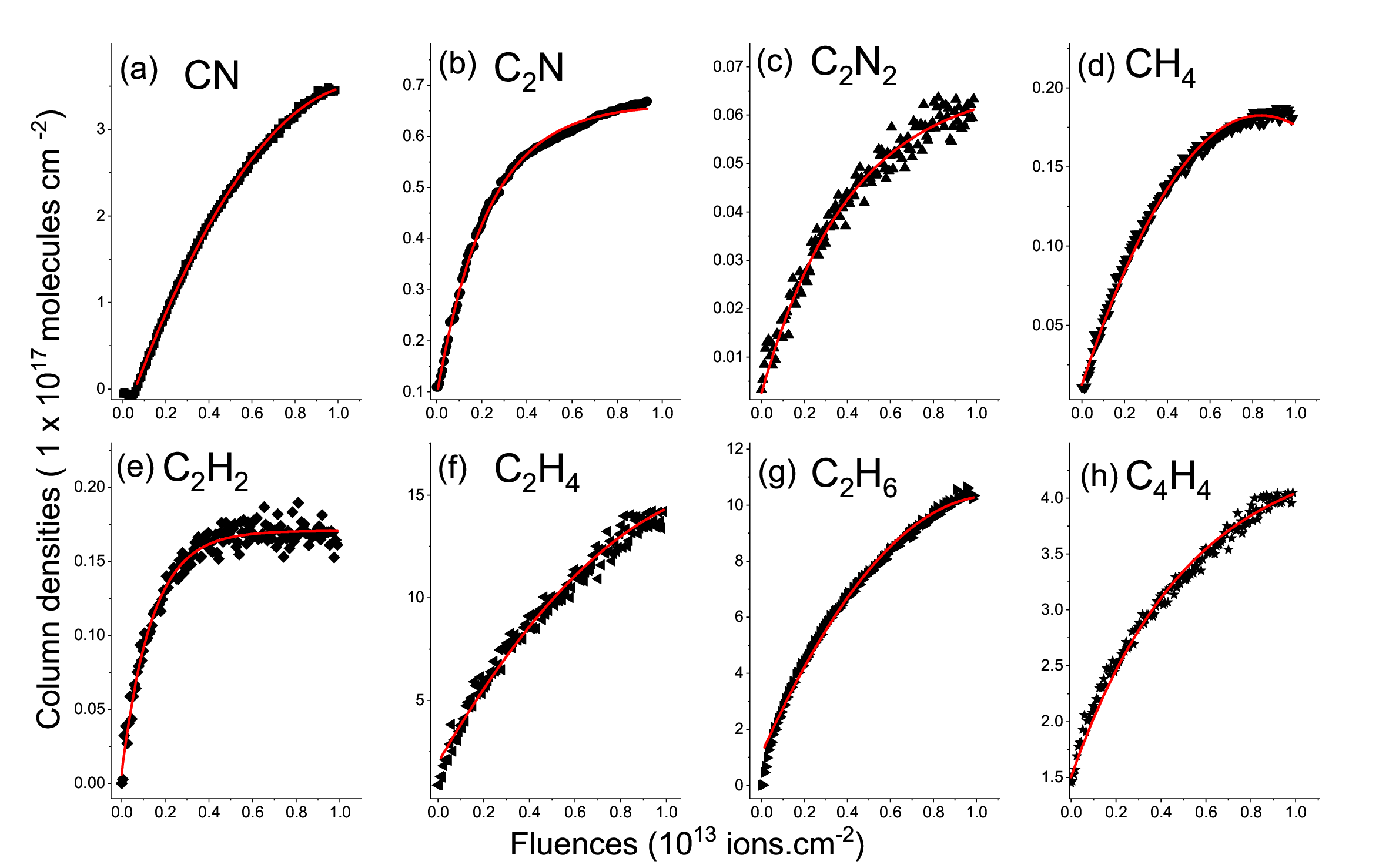}
\caption{Evolution of the column densities of CN-bearing radicals, cyanogen, and hydrocarbon products formed during the irradiation of the PCN:N$_2$ ice as a function of ion fluence. The solid curves represent the best fits obtained using Eq. (\ref{eq4}), which yields the formation ($\sigma_f$) and effective destruction ($\sigma_d^{\rm eff}$) cross sections.}
\label{fig4}
\end{figure*}


\subsection{Destruction of precursor species}

Figure~\ref{fig3}a shows the integrated absorbance areas of the main infrared bands assigned to PCN as a function of ion fluence, including the CH stretching (2998, 2971, 2957, and 2895 cm$^{-1}$), C$\equiv$N stretching (2247 cm$^{-1}$), and CH deformation, rocking, wagging, and skeletal vibration modes at lower wavenumbers. The destruction of PCN was monitored through the isolated C$\equiv$N stretching mode at 2247 cm$^{-1}$, while molecular nitrogen was monitored by the weak matrix-induced absorption centered at 2327 cm$^{-1}$.  The evolution of the two precursor species is presented in Figs.~\ref{fig3}c and \ref{fig3}d, respectively.  The experimental data were fitted using the first-order kinetic expression

\begin{equation}
N(F)=N_0\,\exp(-\sigma_dF),
\label{eq:destruction}
\end{equation}

\noindent where $N_0$ is the initial column density, $F$ is the ion fluence, and $\sigma_d$ is the destruction cross sections.

The exponential model reproduces the experimental data well throughout the investigated fluence range. The best fits yield apparent destruction cross sections of $\sigma_d(\mathrm{PCN}) = 1.85 \times 10^{-13}$ cm$^2$ and $\sigma_d(\mathrm{N}_2) = 2.42 \times 10^{-13}$ cm$^2$. Although the matrix-induced N$_2$ band exhibits a slightly larger apparent disappearance cross section than the PCN band, this result should not be interpreted as evidence that molecular nitrogen is destroyed more efficiently than PCN. Unlike the intrinsic infrared bands of PCN, the weak absorption of solid N$_2$ near 2327 cm$^{-1}$ is activated by interactions within the mixed PCN:N$_2$ ices, and its intensity depends strongly on the local molecular environment. Therefore, the decrease in this band during irradiation reflects not only dissociation of N$_2$ molecules but also radiation-induced structural reorganization of the mixed ice, including changes in the local symmetry and matrix environment surrounding N$_2$.

\subsection{Kinetic analysis of daughter species}

The evolution of the column densities of the daughter species formed during irradiation is presented in Fig.~\ref{fig3}b. The products are grouped according to their chemical families, illustrating the simultaneous formation of nitriles, nitrogen-bearing molecules, hydrocarbons, and CN-bearing radicals as the precursor is progressively destroyed.

The formation kinetics of the daughter species were analyzed by normalizing the product column densities to the initial column density of PCN, $N_0(\mathrm{PCN})$. Although both PCN and N$_2$ are processed during irradiation, PCN constitutes the carbon-bearing precursor from which the observed organic products are formed, whereas molecular nitrogen mainly acts as the surrounding matrix and as a source of reactive nitrogen atoms. Consequently, $N_0(\mathrm{PCN})$ was adopted as the reference column density for the kinetic analysis of the daughter species.
The column density of a daughter species $j$ is described by

\begin{equation}
\frac{N_j(F)}{N_0(\mathrm{PCN})}
\approx
\sigma_{f,j}
\left[
F-\frac{1}{2}\sigma^{\mathrm{eff}}_{d,j}F^2
\right],
\label{eq4}
\end{equation}

\noindent where $N_j(F)$ is the column density of species $j$ at fluence $F$, $\sigma_{f,j}$ is the formation cross section, and $\sigma^{\mathrm{eff}}_{d,j}$ is the effective destruction cross section. Within this approximation, the initial slope of the growth curve is directly proportional to $\sigma_f$, whereas the curvature reflects the competition between product formation and destruction. 

The effective destruction cross section is expressed as
\begin{equation}
\sigma^{\mathrm{eff}}_{d,j}
=
\sigma_{d,\mathrm{PCN}}
+
\sigma_{d,j},
\label{eq5}
\end{equation}

\noindent where $\sigma_{d,\mathrm{PCN}}$ and $\sigma_{d,j}$ are the destruction cross sections of the precursor and daughter species, respectively. Therefore, $\sigma^{\mathrm{eff}}_{d,j}$ accounts for both the depletion of the precursor and the destruction of the newly formed molecule during irradiation.

\begin{figure*}[t]
\centering
\includegraphics[width=\textwidth]{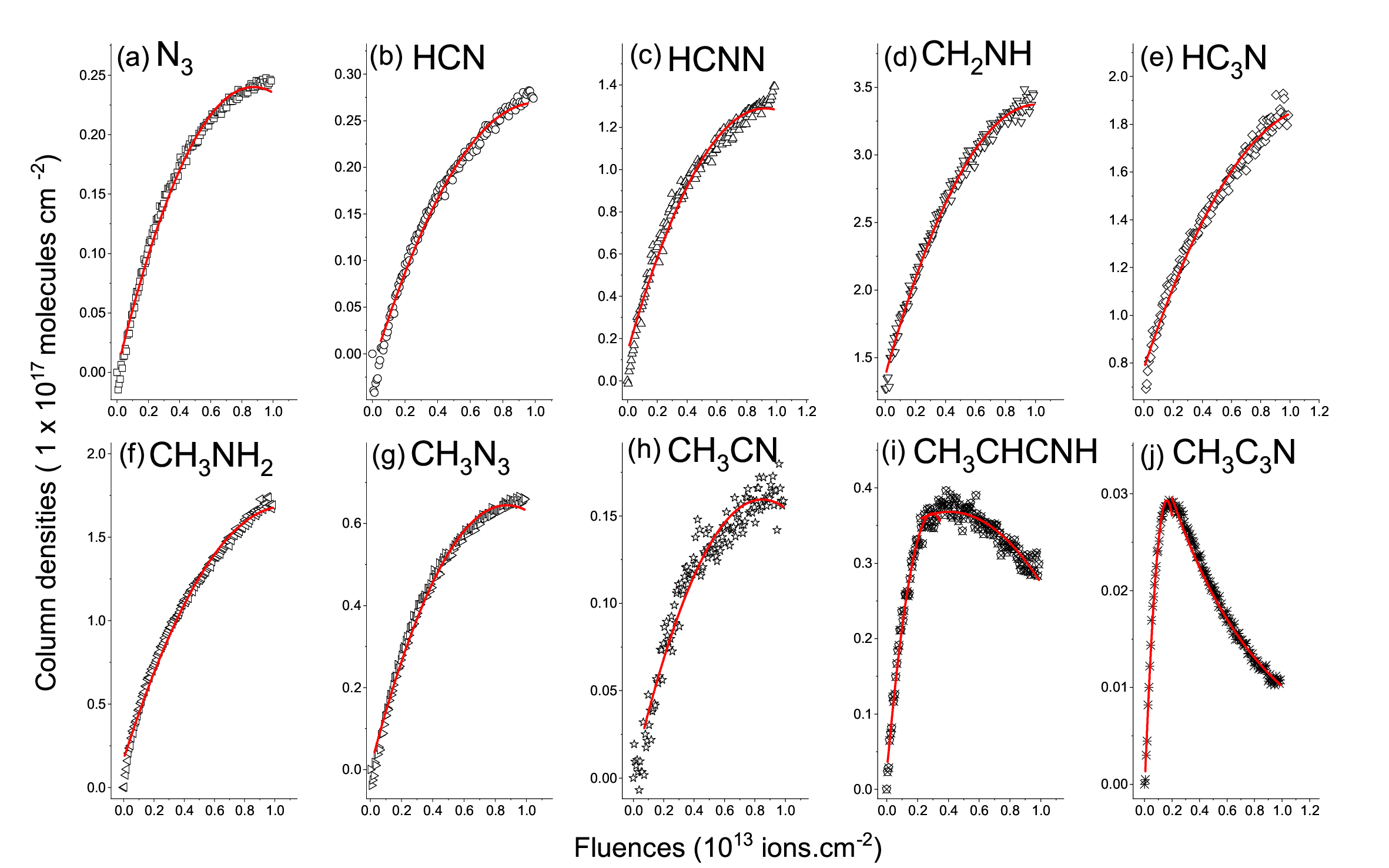}
\caption{Evolution of the column densities of the principal nitrogen-bearing molecules, nitriles, and isonitriles produced during irradiation of the PCN:N$_2$ ice as a function of ion fluence. Solid curves correspond to the fits obtained using Eq. (\ref{eq4}), which yields the formation ($\sigma_f$) and effective destruction ($\sigma_d^{\rm eff}$) cross sections.}
\label{fig5}
\end{figure*}

The experimental growth curves shown in Figs.~\ref{fig4} and \ref{fig5} were fitted using Eq.~(\ref{eq4}), yielding the formation cross section ($\sigma_f$) and the effective destruction cross section ($\sigma_d^{\mathrm{eff}}$) for each quantified product. These parameters, together with the corresponding radiation chemical yields ($G_f$ and $G_d$), are summarized in Table~\ref{tab:table2}. Although the kinetic analysis is normalized to the initial column density of PCN, the observed daughter species are not derived exclusively from the precursor. Ion-induced excitation and partial dissociation of the surrounding N$_2$ matrix also generate reactive nitrogen atoms and radicals, which participate in the reaction network and become incorporated into several of the newly formed molecules. Consequently, the reported formation cross sections represent the overall efficiency of product formation in the irradiated PCN:N$_2$ ice rather than reaction pathways involving only PCN.

The derived formation cross sections span nearly two orders of magnitude, demonstrating that the different reaction pathways do not contribute equally to the radiation chemistry of the PCN:N$_2$ ice. Products exhibiting large $\sigma_f$ values correspond to efficient reaction channels that rapidly convert the primary fragments into stable molecules, whereas smaller formation cross sections indicate less favorable or multistep pathways. Likewise, the effective destruction cross section provides a measure of the stability of each daughter species under continued irradiation, reflecting the dynamic balance between molecular formation and radiolytic destruction. To facilitate the discussion, the identified products were classified into three chemical families according to their molecular structure: (i) nitriles and isonitriles, (ii) nitrogen-bearing species, and (iii) hydrocarbons.

\subsubsection{Nitriles and isonitriles}

Nitriles constitute the largest family of products identified after irradiation, demonstrating that the C$\equiv$N functional group survives energetic processing despite the extensive fragmentation of PCN. Rather than being completely destroyed, the precursor is converted into a chemically diverse set of nitriles, isonitriles, and CN-bearing radicals through dissociation, hydrogen abstraction, molecular rearrangement, and radical recombination reactions \citep{Hudson2004,Abdulgalil2013,Danger2013,Borengasser2026RadiolysisChemistry}.

The kinetic parameters listed in Table~\ref{tab:table2} reveal that the CN radical also exhibits one of the largest formation cross sections among the CN-bearing products, indicating that cleavage of the precursor molecules efficiently generates highly reactive intermediates. These radicals subsequently react with carbon-containing fragments, giving rise to stable nitriles and promoting molecular growth. The detection of CN together with C$_2$N and cyanogen (NCCN/C$_2$N$_2$) implies that radical production and recombination proceed simultaneously throughout the irradiation experiment. 

Among the stable products, CH$_2$CHCN exhibits one of the largest formation cross sections, indicating that dehydrogenation of PCN represents an efficient reaction pathway under heavy-ion irradiation. In contrast to complete molecular fragmentation, this process preserves the carbon skeleton while increasing molecular unsaturation. Similar behavior has been reported during ultraviolet photolysis and ion irradiation of nitrile-containing astrophysical ice analogs, where unsaturated nitriles become increasingly abundant as energetic processing proceeds \citep{Danger2013,Toumi2014PhotolysisStudy,Toumi2016,Borengasser2026RadiolysisChemistry}.

The formation of HC$_3$N likely proceeds through more than one reaction pathway. In addition to carbon-chain growth involving reactions between CN radicals and unsaturated hydrocarbon fragments, the efficient production of acrylonitrile (CH$_2$CHCN) observed in the present work suggests that successive dehydrogenation following H-atom abstraction from PCN provides a direct route toward HC$_3$N \citep{Hudson2004,Abdulgalil2013,Coupeaud2006}. The two mechanisms are expected to operate simultaneously under swift heavy-ion irradiation, contributing to the observed cyanoacetylene abundance.

The simultaneous observation of HC$_3$N, CH$_3$CN, HCNN, CH$_3$C$_3$N, and ketenimine-like species suggests that several competing reaction channels preserve the nitrile functionality while redistributing carbon and nitrogen atoms into new molecular structures. In contrast to the monotonic growth observed for most nitriles,
Fig.~\ref{fig5} shows that CH$_3$C$_3$N reaches a maximum
column density at relatively low fluence and subsequently decreases. This behavior indicates that the radiolytic destruction of CH$_3$C$_3$N becomes dominant over its formation at higher fluences, suggesting that methylcyanoacetylene is a transient intermediate efficiently consumed by secondary dissociation and radical-driven reactions. These results further support energetic processing as an efficient pathway for the formation of increasingly complex nitriles in nitrogen-rich astrophysical ices.

\begin{table*}[t]
\centering
\scriptsize
\caption{Radiolysis products quantified in irradiated PCN:N$_2$ ice at 10 K.
}

\label{tab:table2}

\setlength{\tabcolsep}{3pt}
\renewcommand{\arraystretch}{1.05}

\begin{tabular}{p{1.4cm} p{2.6cm} p{1.2cm} p{1.6cm} p{1.7cm} p{1.7cm} p{1.0cm} p{1.0cm}}
\hline
Species &
Common name &
Band used &
$A_v$ &
$\sigma_f$ (cm$^{2}$) &
$\sigma_d^{\rm eff}$ (cm$^{2}$) &
$G_f$ &
$G_d^{\rm eff}$\\
&
&
(cm$^{-1}$) &
&
&
&
&
\\
\hline

\multicolumn{8}{c}{\textbf{(i) Nitriles, isonitriles and CN-bearing radicals}}\\
\hline

HCN &
Hydrogen cyanide &
827$^{(a)}$ &
$1.1\times10^{-18}$ &
$1.05\times10^{-14}$ &
$9.93\times10^{-14}$ &
0.071 &
0.674 \\

HCNN &
HCNN &
1832$^{(b)}$ &
$4.1\times10^{-17}$ &
$4.62\times10^{-14}$ &
$1.08\times10^{-13}$ &
0.314 &
0.733 \\

HC$_3$N &
Cyanoacetylene &
2066$^{(c)}$ &
$1.0\times10^{-17}$ &
$3.39\times10^{-14}$ &
$8.38\times10^{-14}$ &
0.230 &
0.569 \\

CH$_3$CN &
Acetonitrile &
2263$^{(d)}$ &
$2.2\times10^{-18}$ &
$6.83\times10^{-15}$ &
$1.18\times10^{-13}$ &
0.046 &
0.801 \\

CH$_2$CHCN &
Acrylonitrile &
2230$^{(b,e,f)}$ &
$4.0\times10^{-18}$ &
$4.32\times10^{-14}$ &
$3.38\times10^{-13}$ &
0.293 &
2.294 \\

CH$_3$CHCNH$^{\ddagger}$ &
Ketenimine-like species &
2034$^{(f,i)}$ &
$7.2\times10^{-17}$ &
$4.04\times10^{-15}$ &
$2.46\times10^{-13}$ &
0.027 &
1.670 \\

CH$_3$C$_3$N &
Methylcyanoacetylene &
2340$^{(g)}$ &
$8.4\times10^{-18}$ &
$4.72\times10^{-14}$ &
$1.35\times10^{-13}$ &
0.320 &
0.917 \\

NCCN/C$_2$N$_2$ &
Cyanogen &
2167$^{(d,h)}$ &
$2.2\times10^{-18}$ &
$2.08\times10^{-15}$ &
$1.06\times10^{-13}$ &
0.014 &
0.720 \\

CN &
Cyano radical &
2089$^{(b)}$ &
$3.2\times10^{-18}$ &
$1.28\times10^{-13}$ &
$8.83\times10^{-14}$ &
0.869 &
0.599 \\

C$_2$N &
Dicarbon nitride radical &
1054$^{(b)}$ &
$1.1\times10^{-17}$ &
$4.20\times10^{-14}$ &
$1.19\times10^{-13}$ &
0.285 &
0.808 \\

\hline

\multicolumn{8}{c}{\textbf{(ii) Nitrogen-bearing species}}\\
\hline

CH$_2$NH &
Methanimine &
1347$^{(i)}$ &
$7.6\times10^{-18}$ &
$7.33\times10^{-14}$ &
$9.88\times10^{-14}$ &
0.498 &
0.671 \\

CH$_3$NH$_2$ &
Methylamine &
2863$^{(b,j)}$ &
$1.7\times10^{-17}$ &
$5.26\times10^{-14}$ &
$9.40\times10^{-14}$ &
0.357 &
0.638 \\

CH$_3$N$_3$ &
Methyl azide &
1293$^{(i)}$ &
$1.8\times10^{-17}$ &
$2.71\times10^{-14}$ &
$1.14\times10^{-13}$ &
0.184 &
0.774 \\

N$_3^{-}$ &
Azide anion &
1656$^{(b,k)}$ &
$4.0\times10^{-17}$ &
$1.01\times10^{-14}$ &
$1.14\times10^{-13}$ &
0.069 &
0.774 \\

\hline

\multicolumn{8}{c}{\textbf{(iii) Hydrocarbons}}\\
\hline

CH$_4$ &
Methane &
1305$^{(c,l)}$ &
$6.4\times10^{-18}$ &
$7.53\times10^{-15}$ &
$1.19\times10^{-13}$ &
0.051 &
0.808 \\

C$_2$H$_2$ &
Acetylene &
3227$^{(m,n,o)}$ &
$3.6\times10^{-17}$ &
$6.84\times10^{-15}$ &
$1.50\times10^{-13}$ &
0.046 &
1.018 \\

C$_2$H$_4$&
Ethylene &
949.5$^{(p)}$ &
$8.4\times10^{-18}$ &
$1.59\times10^{-13}$ &
$7.21\times10^{-14}$ &
2.437 &
0.489 \\

C$_2$H$_6$ &
Ethane &
2949$^{(m,p)}$ &
$2.1\times10^{-18}$ &
$1.16\times10^{-13}$ &
$9.27\times10^{-14}$ &
2.145 &
0.629 \\

C$_4$H$_4$ &
Vinylacetylene &
3282$^{(n,o,q)}$ &
$1.3\times10^{-17}$ &
$7.47\times10^{-14}$ &
$8.38\times10^{-14}$ &
0.507 &
0.569 \\

\hline
\end{tabular}

\vspace{0.15cm}

\tablefoot{Species are grouped according to the chemical classes shown in
Fig.~\ref{fig6}. Adopted band strength ($A_v$, cm molecule$^{-1}$), formation cross
section ($\sigma_f$), effective destruction cross section
($\sigma_d^{\rm eff}$), and radiation chemical yields ($G_f$ and
$G_d^{\rm eff}$, molecules per 100 eV). Superscripts identify the literature source used for the infrared band assignment and/or the adopted band strength.  The symbol $^{\ddagger}$ denotes a tentative assignment to ketenimine-like species (CH$_3$CHCNH/H$_2$CCNH). References:
$^{(a)}$ \cite{Gerakines2022};
$^{(b)}$ Calculated by DFT \cite{Wu2013};
$^{(c)}$ \cite{Abdulgalil2013};
$^{(d)}$ \cite{Danger2013};
$^{(e)}$ \cite{Hudson2004};
$^{(f)}$ \cite{Borengasser2026RadiolysisChemistry};
$^{(g)}$ Calculated from the integrated absorption coefficient reported by \cite{Del96};
$^{(h)}$ \cite{Moore2010};
$^{(i)}$ \cite{quinto2011};
$^{(j)}$ \cite{theule2011};
$^{(k)}$ \cite{Jamieson2007};
$^{(l)}$ \cite{dHendecourt1986};
$^{(m)}$ \cite{Chuang2021};
$^{(n)}$ \cite{abplanalp2020};
$^{(o)}$ \cite{Lo2020};
$^{(p)}$ \cite{Bohn1994};
$^{(q)}$ \cite{kim09}.
}
\end{table*}


\subsubsection{Nitrogen-bearing species}

Besides nitriles, irradiation of the PCN:N$_2$ ice produces a second family of products composed of nitrogen-bearing molecules, namely CH$_2$NH, CH$_3$NH$_2$, CH$_3$N$_3$, and N$_3^-$. Their formation demonstrates that molecular nitrogen actively participates in the radiation chemistry of the mixed ice instead of acting solely as an inert matrix. Dissociation and electronic excitation of N$_2$ generate reactive nitrogen atoms that subsequently react with carbon-containing radicals produced from PCN fragmentation.

Among these products, CH$_2$NH exhibits the largest formation cross section, indicating that imine formation constitutes one of the most efficient nitrogen incorporation pathways under the present experimental conditions \citep{quinto2011,Wu2013}. Methylamine (CH$_3$NH$_2$) has attracted considerable astrochemical interest because it is considered a potential precursor of more complex nitrogen-containing organic molecules. Its somewhat smaller formation cross section suggests that hydrogenation reactions compete with dehydrogenation and radical recombination during irradiation.

The formation of CH$_3$N$_3$ and N$_3^-$ further illustrates the rich nitrogen chemistry promoted by heavy-ions in N$_2$-rich ices. Although these species are produced with comparatively lower efficiencies, their detection confirms that nitrogen atoms released from the matrix are incorporated into several distinct functional groups. Similar nitrogen-bearing products have been reported in irradiated N$_2$-containing astrophysical ice analogs, supporting the interpretation that energetic processing efficiently redistributes nitrogen among multiple reaction channels \citep{Jamieson2007,Hudson2002TheClouds,Wu2013}.

\subsubsection{Hydrocarbons}

Hydrocarbons constitute the third major family of daughter species identified after irradiation, indicating that fragmentation of the PCN carbon backbone is followed by efficient radical recombination. The observed products include CH$_4$, C$_2$H$_2$, C$_2$H$_4$, C$_2$H$_6$, and C$_4$H$_4$, covering both saturated and unsaturated hydrocarbons.

The kinetic parameters show that C$_2$H$_4$ and C$_2$H$_6$ exhibit the largest formation cross sections among the hydrocarbons. This behavior suggests that C$_2$ fragments generated during cleavage of the ethyl group constitute stable intermediates during irradiation. The simultaneous production of ethylene and ethane suggests that hydrogen abstraction and hydrogen addition occur simultaneously within different local environments along the ion track

Methane and acetylene are formed with comparatively smaller formation cross sections and probably represent terminal products of extensive molecular fragmentation. In contrast, the detection of C$_4$H$_4$ indicates that radical recombination also promotes carbon-chain growth. Similar hydrocarbon distributions have been reported for energetic processing of hydrocarbon-rich astrophysical ices, where fragmentation and recombination occur simultaneously under nonequilibrium conditions \citep{Abplanalp2016RADIATION,abplanalp2020,Bohn1994}.


\begin{figure*}
    \centering
 \includegraphics[width=0.80\linewidth]{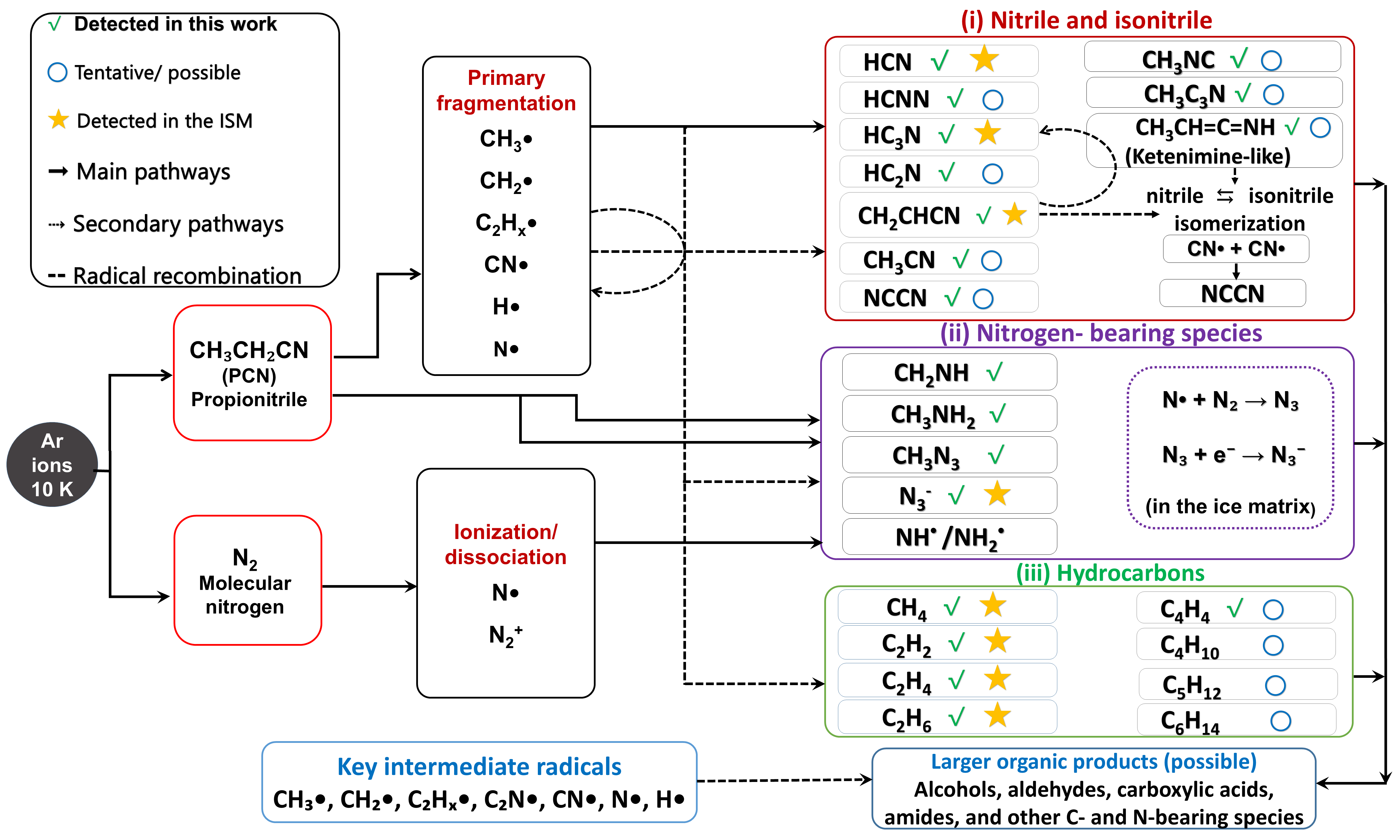}
 \caption{Proposed reaction network for the radiolysis of PCN:N$_2$ ice by 40 MeV $^{40}$Ar$^{9+}$ ions at 10 K. The scheme summarizes the main fragmentation, radical recombination, isomerization, and nitrogen incorporation pathways inferred from the infrared identifications and kinetic analysis. The radicals shown in the "Primary fragmentation" box are representative examples of how reactive intermediates are inferred from the experimental results and do not constitute an exhaustive list of all fragments produced during irradiation.}
    \label{fig6}
\end{figure*}

\subsection{Reaction pathways and astrochemical implications}

The reaction network proposed based on the present experimental results is summarized in Fig.~\ref{fig6}. The scheme combines the infrared identifications with the kinetic parameters derived from the growth curves, providing a comprehensive picture of the chemical evolution of irradiated PCN:N$_2$ ice. Radiolysis is initiated by electronic excitation and ionization induced by the 40 MeV $^{40}$Ar$^{9+}$ ions, leading to the dissociation of PCN and partial activation of the surrounding N$_2$ matrix. These primary processes generate reactive fragments, including CN, CH$_x$, C$_2$H$_x$, nitrogen atoms, and nitrogen-containing radicals, which subsequently undergo radical recombination, hydrogen abstraction, dehydrogenation, and molecular rearrangement reactions.

As illustrated in Fig.~\ref{fig6}, the chemistry can be grouped into three interconnected pathways. The first preserves the C$\equiv$N functional group, producing nitriles and isonitriles such as HCN, HC$_3$N, CH$_3$CN, CH$_2$CHCN, HCNN, and cyanogen (NCCN/C$_2$N$_2$). The second incorporates reactive nitrogen atoms released from the N$_2$ matrix into newly formed molecules, yielding CH$_2$NH, CH$_3$NH$_2$, CH$_3$N$_3$, and N$_3^-$. The third involves fragmentation of PCN followed by recombination of hydrocarbon radicals, producing CH$_4$, C$_2$H$_2$, C$_2$H$_4$, C$_2$H$_6$, and C$_4$H$_4$. The formation cross sections listed in Table~\ref{tab:table2} demonstrate that these reaction channels operate simultaneously throughout irradiation, continuously recycling the atoms released from the precursor into a chemically diverse inventory of daughter species.

Several of the molecules produced in the present experiment, including HCN, HC$_3$N, CH$_3$CN, CH$_2$CHCN, CH$_2$NH, and CH$_3$NH$_2$, have been identified in dense molecular clouds, hot molecular cores, protostellar envelopes, and protoplanetary disks, where they participate in the chemistry leading to increasingly complex organic molecules \citep{Belloche2022,Daly2013,Xue2020,Shingledecker2021,Bergner2022,Gulin2022}. The efficient formation of HC$_3$N and CH$_2$CHCN demonstrates that energetic processing simultaneously promotes carbon-chain growth and preservation of the C$\equiv$N functional group, providing an efficient solid-state route to nitriles commonly observed in the ISM. Likewise, the formation of methanimine and methylamine supports previous suggestions that the energetic processing of nitrogen-rich ices contributes to the production of simple prebiotic precursors in cold astrophysical environments \citep{theule2011,quinto2011,Wu2013,Enrique-Romero2026,Ricca2026,Wang2026}.

The present results also demonstrate that molecular nitrogen is not merely an inert diluting matrix. Partial dissociation and electronic excitation of N$_2$ are expected to generate reactive nitrogen atoms that become incorporated into several chemically distinct products, including CH$_2$NH, CH$_3$NH$_2$, CH$_3$N$_3$, and N$_3^-$. Similar nitrogen chemistry has been reported for irradiated and photolyzed N$_2$-containing astrophysical ice analogs, confirming that energetic processing efficiently converts condensed molecular nitrogen into chemically available reactive species \citep{Jamieson2007,Hudson2002TheClouds,Vasconcelos2017System,Borengasser2026RadiolysisChemistry,Tribbett2026,Zhang2025}.

The formation of acrylonitrile is particularly relevant in the context of Titan and other nitrogen-rich planetary environments, where this molecule has been detected or proposed as a precursor of increasingly complex organic materials \citep{Czaplinski2025}. More generally, the coexistence of nitriles, nitrogen-bearing molecules, and hydrocarbons produced from a single precursor demonstrates the ability of heavy cosmic rays to increase molecular complexity within icy mantles under dense cloud conditions. Unlike previous laboratory studies focused primarily on product identification, the present work provides quantitative destruction and formation cross sections together with radiation chemical yields for the irradiation of a PCN-rich ice by swift heavy-ions. These experimental kinetic parameters provide valuable constraints for astrochemical models that explicitly include solid-state processing by heavy cosmic rays and contribute to a better understanding of the chemical evolution of nitrogen-bearing organic molecules in dense molecular clouds, protostellar environments, protoplanetary disks, Titan, and other nitrogen-rich bodies of the Solar System \citep{Hudson2004,Abdulgalil2013,Danger2013,Bergner2019Disks}.

\section{Conclusions}

The radiolysis of a PCN:N$_2$ (1:10) ice mixture at 10 K by 40 MeV $^{40}$Ar$^{9+}$ ions was investigated by in situ FTIR spectroscopy to evaluate the chemical evolution of PCN under conditions relevant to heavy cosmic-ray processing of nitrogen-rich astrophysical ices. The main conclusions are summarized as follows:

\begin{enumerate}

\item The infrared spectrum of the pristine PCN:N$_2$ ice shows that dilution in molecular nitrogen preserves the vibrational fingerprint of PCN, with only minor matrix-induced shifts and broadening. The C$\equiv$N stretching band at 2247 cm$^{-1}$ remained suitable for determining the precursor column density throughout the irradiation experiment.

\item Heavy-ion irradiation efficiently destroys PCN and simultaneously produces a chemically diverse inventory of daughter species. Nineteen radiolysis products were quantified, including nitriles, isonitriles, nitrogen-bearing molecules, hydrocarbons, and reactive radicals, demonstrating the extensive molecular reorganization induced by energetic processing.

\item Formation and effective destruction cross sections were determined for the major daughter species together with their radiation chemical yields. The kinetic analysis shows that several reaction channels operate simultaneously, with efficient preservation of the C$\equiv$N functional group, incorporation of nitrogen released from the N$_2$ matrix, and fragmentation followed by hydrocarbon formation.

\item The reaction network proposed based on the experimental results indicates that heavy-ion irradiation continuously recycles the carbon and nitrogen atoms initially present in the precursor into increasingly complex organic molecules. The formation of species such as HC$_3$N, CH$_2$CHCN, CH$_2$NH, and CH$_3$NH$_2$ demonstrates that energetic processing provides an efficient solid-state route for producing molecules widely detected in dense molecular clouds and other nitrogen-rich astrophysical environments.

\item The quantitative kinetic parameters reported here, including destruction and formation cross sections together with radiation chemical yields, provide new experimental constraints for astrochemical models that incorporate solid-state chemistry induced by heavy cosmic rays. This work presents one of the first systematic kinetic investigations of the radiolysis of PCN diluted in molecular nitrogen under swift heavy-ion irradiation, providing a benchmark for understanding the chemical evolution of nitrile-rich astrophysical ices.

\end{enumerate}

\begin{acknowledgements}
A. L. F. de Barros acknowledges the support of the GANIL Long-Term Visitor Program. She also acknowledges financial support from the Brazilian funding agencies Conselho Nacional de Desenvolvimento Científico e Tecnológico (CNPq; Productivity Fellowship No. 301535/2026-4), Fundação de Amparo à Pesquisa do Estado do Rio de Janeiro (FAPERJ; Grants E-26/204.934/2024, E-26/210.836/2025, E-26/200.188/2026 and E-26/203.099/2026), Financiadora de Estudos e Projetos (FINEP; Grant No. 01.25.0309.01, Ref. 1455/24), and Coordenação de Aperfeiçoamento de Pessoal de Nível Superior (CAPES; Finance Code 001). The experiment was performed at Grand Accélérateur National d’Ions Lourds (GANIL) by means of CIRIL Interdisciplinary Platform, part of CIMAP laboratory, Caen, France. D. Dubois acknowledges support from the French Agence Nationale de la Recherche. R. Sreeja acknowledges Normandy Region for Financial support. We acknowledge support from the French Agence Nationale de la Recherche ANR MIRRPLA project of PEPR Origins (France 2030 investment plan, reference ANR-22-EXOR-0012). 
\end{acknowledgements}

\appendix

\section{Additional experimental parameters}

\begin{table*}[ht]
\centering
\scriptsize
\caption{Observed infrared bands in irradiated PCN:N$_2$ ice at 10 K and tentative assignments based on the bands marked in Fig.~\ref{fig2}.}
\label{tab:table1}

\setlength{\tabcolsep}{3pt}
\renewcommand{\arraystretch}{0.95}

\begin{tabular}{cccccc}
\hline
Fig.\ref{fig2}  &
Observed band &
Integration region &
Possible species &
Vibrational assignment &
Observation \\
panel&
(cm$^{-1}$) &
(cm$^{-1}$) &
&
&
\\
\hline

(a) & 3282 & 3192--3341 &
CH$_3$NH$_2^{a,b}$, C$_2$H$_2^{c}$, C$_4$H$_4^{d,e,f,j,k}$ &
$\equiv$C--H / N--H stretching &
New band/tentative assignment \\

(a) & 3254 & 3192--3341 &
CH$_3$CH$_2$CN$^{g,h}$ &
$\nu_4$ + $\nu_{10}$ CH$_3$ asym. stretching &
PCN precursor / overlap \\

(a) & 3227 & 3192--3341 &
C$_2$H$_2^{e,i,j,k}$ &
$\equiv$C--H stretching &
New band \\

(a) & 3138 & 3019--3191 &
HC$_3$N$^{l}$,CH$_2$CHCN$^{m}$ & 
=C--H stretching &
New band \\

(a) & 3085 & 3019--3191 &
CH$_3$CH$_2$CN$^{g,h}$ &
$\nu_4$ + $\nu_{11}$ CH$_3$ asym. stretching &
PCN precursor / overlap \\

\hline

(b) & 2998 & 2965--3020 &
CH$_3$CH$_2$CN$^{l,n,g,h}$ &
$\nu_1$ CH$_3$ asym. stretching &
PCN precursor \\

(b) & 2971 & 2931--2965 &
CH$_3$CH$_2$CN$^{l,n,g}$, C$_2$H$_6^{
o,aj}$ &
$\nu_{14}$ CH$_3$ asym. stretching &
PCN precursor \\

(b) & 2957 & 2931--2965 &
CH$_3$CH$_2$CN$^{l,n,g}$ &
$\nu_{15}$ CH$_2$ asym. stretching&
PCN precursor \\

(b) & 2949 & 2940--2960 &
C$_2$H$_6$$^{c,e,aj,ak}$ &
CH asym. stretching&
New band \\
\hline

(c) & 2922 & 2868--2929 &
CH$_3$CH$_2$CN$^{l,g}$ &
$\nu_{2}$ CH$_3$ symmetric stretching &
PCN precursor / overlap \\

(c) & 2895 & 2868--2909 &
CH$_3$CH$_2$CN$^{l,n,g}$, C$_2$H$_6^{c,e}$ &
$\nu_2$ CH$_3$ sym. stret / $\nu_3$ CH$_2$ &
PCN precursor / overlap \\

(c) & 2864 & 2860--2890 &
C$_4$H$_{10}^{c}$, C$_5$H$_{12}^{c}$, CH$_3$NH$_{2}^{p}$ &
C--H stretching &
New band \\

\hline

(d) & 2763 & 2550--2860 &
CH$_4$ in an N$_2$ matrix$^{l,r,s}$ &
C--H combination/overtone modes &
New band \\

(d,e) & 2340 & 2331--2352 &
CH$_3$C$_3$N$^{l}$, NCCN/C$_2$N$_2$$^{t}$ &
2$\nu_5$ overtone (C$\equiv$N-related) &
New band \\

\hline

(e) & 2327 & 2321--2335 &
N$_2^{c,v}$ &
Matrix-induced N$_2$ absorption &
N$_2$ precursor \\

(e) & 2264 & 2237--2288 &
HC$_3$N$^{l,u,r,s,w,al}$, CH$_3$CN$^{l,t,u,x,y}$ &
overlapping $\nu$(C$\equiv$N)  &
New band \\

(e) & 2247 & 2237--2288 &
CH$_3$CH$_2$CN$^{l,n,g,t}$ &
$\nu_4$ C$\equiv$N stretching &
PCN precursor \\

(e) & 2230 & 2192--2239 &
CH$_2$CHCN$^{z,aa,w,ab}$ &
$\nu$(C$\equiv$N) &
New band \\

\hline

(f) & 2167 & 2160--2180 &
CH$_3$NC$^{t,u,m,x,ab}$, CH$_3$CH$_2$NC$^{m}$, NCCN/C$_2$N$_2^{l,t,y,al}$ &
$\nu$(N$\equiv$C) &
New band \\

(f) & 2140 & 2135--2150 &
NCCN/C$_2$N$_2^{p}$, (CH$_3$)$_3$CNC$^{t}$ &
C$\equiv$N-related vibration &
New band \\

(f) & 2132 & 2120--2145 &
CH$_2$CHNC$^{l,t,r,y}$ &
N$\equiv$C stretch &
New band \\

(f) & 2089 & 2070--2100 &
CN$^{p}$, HCN$^{ad,ae,m,x,y}$, &
$\nu$(C$\equiv$N) &
New band \\

(f) & 2066 & 2045--2085 &
NCNC$^{t,p}$, CN$^{- ak}$, HC$_3$N$^{l,u,r,s}$ &
$\nu$(C$\equiv$N) &
New band \\

(f) & 2034 & 1980--2044 &
CH$_3$CHCNH$^{m,x,aa}$, H$_2$CCNH$^{t,u,m,x,y}$,CH$_2$CNH$^{ac}$ &
CCN vibration &
New band \\

\hline

(g) & 1832 & 1790--1846 &
HCNN$^{p}$ &
Combination/overtone band &
New band \\

(g) & 1725 & 1698--1740 &
Unidentified & C=N/C=C-related vibration &
New band \\

(g) & 1656 & 1633--1676 &
CH$_2$NH$^{ak,ag}$, N$_3^{- p,ah}$ &
C=N stretching / $\nu_3$(N$_3^-$) &
New band \\

(g) & 1643 & 1633--1676 &
CH$_2$NH$^{ab,ag}$ &
C=N-related vibration &
New band \\

(g) & 1627  & 1612--1632 &
CH$_2$CHCN$^{r}$ &
C=C stretch &
New band \\

(g) & 1602 & 1580--1620 &
CH$_2$CHCN$^{l,w}$, C$_4$H$_4^{d,f,j,ai}$ &
C=N stretching / $\nu_5$ &
New band \\

\hline

(h) & 1468 & 1448--1481 &
CH$_3$CH$_2$CN$^{l,n,g,h}$ &
$\nu_5$ CH$_3$ deformation &
PCN precursor / overlap \\

(h) & 1462 & 1448--1481 &
CH$_3$CH$_2$CN$^{l,n,g}$, C$_2$H$_6^{c,o,e,aj}$ &
$\nu_{16}$ CH$_3$ deformation &
PCN precursor / overlap \\

(h) & 1435 & 1397--1448 &
CH$_3$CH$_2$CN$^{l,n,g,h}$, C$_2$H$_4^{c,o,e,k}$ &
$\nu_6$ CH$_2$ deformation &
PCN precursor / overlap \\

(h) & 1429 & 1397--1448 &
CH$_3$CH$_2$CN$^{l,n,g}$ &
Matrix-split deformation component &
PCN precursor / overlap \\

(h) & 1384 & 1372--1396 &
CH$_3$CH$_2$CN$^{l,n,g}$ &
$\nu_7$ CH$_3$ bending &
PCN precursor \\

(h) & 1348 & 1320--1360 &
CH$_2$NH$^{ag}$ &
$\nu_6$ HCNH deformation  &
New band \\

(h) & 1325 & 1309--1340 &
CH$_3$CH$_2$CN$^{n,g}$ &
$\nu_8$ CH$_2$ wagging &
PCN precursor \\

\hline

(h,i) & 1305 & 1290--1310 &
CH$_4^{u,m,x,af,ab}$ &
CH deformation &
New band \\

(i) & 1293 & 1290--1310 &
C$_4$H$_{10}^{c}$, CH$_3$N$_3$$^{ag}$ &
CH deformation &
New band / overlap \\

(i) & 1283 & 1270--1290 &
CH$_2$CHCN$^{w}$ &
CH deformation &
New band \\

(i) & 1268 & 1260--1280 &
CH$_3$CH$_2$CN$^{l,n,g}$ &
$\nu_{17}$ CH$_2$ wagging &
PCN precursor \\

(i) & 1256 & 1245--1265 &
C$_4$H$_4^{e,f}$ &
Combination mode &
New band \\

(i,j) & 1098 & 1090--1106 &
CH$_3$CH$_2$CN$^{l,n,g}$ &
$\nu_{18}$ CH$_2$ rocking  &
PCN precursor \\

\hline

(j) & 1077 & 1066--1090 &
CH$_3$CH$_2$CN$^{l,n,g}$ &
C--N stretching &
PCN precursor \\

(j) & 1054 & 1049--1065 &
 C$_2$N$^{p}$ &
C--N stretching &
New band \\

(j) & 1042 & 1030--1055 &
CH$_3$NH$_2^{a,b}$, N$_3^{- p}$ &
C--N stretching / $\nu_3$(N$_3^-$) &
New band \\

(j) & 1017 & 1010--1025 &
CH$_3$CH$_2$CN$^{l,g}$ &
$\nu_{10}$ C--C stretching &
PCN precursor \\

(j) & 1007 & 998--1010 &
 CH$_3$CH$_2$CN$^{l,n,g}$ &
 C--C stretching &
PCN precursor \\

(j) & 950 & 880--1000 &
C$_2$H$_4^{c,o,e,aj}$ &
CH deformation &
New band \\

\hline

(k) & 873 & 850--891 &
HCNN$^{p,q}$, C$_4$H$_4^{e,f}$,CH$_2$CNH$^{ac}$, CH$_2$CHCN$^{l,w,ab}$ &
C--C stretching / bending modes &
New band \\

(k) & 836 & 812--858 &
CH$_3$CH$_2$CN$^{l,n,g}$ &
$\nu_{11}$ C--C stretching &
PCN precursor \\

(k) & 827 & 780--860 &
HCN$^{l,ad}$, C$_2$H$_6^{c,o,aj,ak}$ &
CH bending &
New band \\

(k) & 783 & 775--811 &
CH$_3$CH$_2$CN$^{l,n,g}$ &
$\nu_{19}$ CH$_2$ rocking &
PCN precursor \\

(k) & 765 & 760--790 &
CH$_3$CH$_2$CN$^{l,g}$, HC$_3$N$^{l}$ &
CH$_2$ rocking / skeletal mode &
PCN precursor / overlap \\

(k) & 761 & 729--796 &
NCCN/C$_2$N$_2^{l,ad}$, C$_2$H$_2^{e,i}$ &
$\nu_4+\nu_5$ &
New band \\

(k) & 686 & 670--700 &
CH$_2$CHCN$^{l,w,r}$,CH$_2$CNH$^{ac}$ &
$\nu_{14}$ torsion &
New band \\

\hline
\end{tabular}

\vspace{0.15cm}

\tablefoot{Multiple assignments are listed only when overlapping contributions at the same wavenumber region cannot be excluded. Superscripts indicate the literature sources used for each assignment.  References:
$^{a}$\cite{theule2011}
$^{b}$\cite{Bossa2008}
$^{c}$\cite{Bohn1994}
$^{d}$\cite{abplanalp2020}
$^{e}$\cite{Chuang2021}
$^{f}$\cite{kim09}
$^{g}$\cite{Heise1981}
$^{h}$\cite{nnamvondo2019}
$^{i}$\cite{Knez2012}
$^{j}$\cite{Cuylle2014}
$^{k}$\cite{Pereira2020IonSimulations}
$^{l}$\cite{Del96}
$^{m}$\cite{Hager2025}
$^{n}$\cite{Hudson2020}
$^{o}$\cite{ben06}
$^{p}$\cite{Wu2013}
$^{q}$\cite{Ogilvie1968}
$^{r}$\cite{Toumi2014PhotolysisStudy}
$^{s}$\cite{Coupeaud2006}
$^{t}$\cite{Hudson2004}
$^{u}$\cite{Abdulgalil2013}
$^{v}$\cite{Bernstein1999}
$^{w}$\cite{Toumi2016}
$^{x}$\cite{Borengasser2026RadiolysisChemistry}
$^{y}$\cite{Danger2013}
$^{z}$\cite{Bernstein1997}
$^{aa}$\cite{Krim2019}
$^{ab}$\cite{Danger2011HCN,Danger2011}
$^{ac}$\cite{Kameneva2017}
$^{ad}$\cite{Moore2010}
$^{ae}$\cite{Jamieson2009AIces}
$^{af}$\cite{dHendecourt1986}
$^{ag}$\cite{quinto2011}
$^{ah}$\cite{Jamieson2007}
$^{ai}$\cite{Lo2020}
$^{aj}$\cite{Kim2010LaboratoryChemistry}
$^{ak}$\cite{Kim2011}
$^{al}$\cite{Ennis2018}.
}
\end{table*}

\bibliography{acs-achemso}

\end{document}